\documentclass[
]{ceurart}

\usepackage{listings}
\usepackage{booktabs}
\usepackage{multirow}
\usepackage{tabularx}
\usepackage{array}
\usepackage{subcaption}
\usepackage{balance}
\usepackage[normalem]{ulem}

\begin{document}

\copyrightyear{2026}
\copyrightclause{Copyright for this paper by its authors.
  Use permitted under Creative Commons License Attribution 4.0
  International (CC BY 4.0).}

\conference{DaQuaMRec Workshop at RecSys '26, September 28, 2026, Minneapolis, Minnesota, USA}

	\title{LUMIR: Decoupled Multimodal Representation Learning \textcolor{black}{from Heterogeneous Data} for Next-POI Recommendation}

%
\author[1,2]{Zehui Wang}[%
orcid=0000-0001-5401-8773,
email=Zehui.Wang@rwu.de,
]
\cormark[1]
\address[1]{University of Applied Sciences Ravensburg-Weingarten, Doggenriedstrasse, Weingarten, Weingarten,  88250, Germany}
\address[2]{University of Klagenfurt, Universitätsstraße 65–67, Klagenfurt am W{\"o}rthersee, 9020, Austria}

\author[1]{Wolfram H{\"o}pken}[%
orcid=0000-0002-4175-1295,
email=Wolfram.Hoepken@rwu.de,
]

\author[2]{Dietmar Jannach}[%
orcid=0000-0002-4698-8507,
email=Dietmar.Jannach@aau.at,
]

\cortext[1]{Corresponding author.}

\begin{abstract}
	As multimodal tourism-related information collected from heterogeneous data sources has become increasingly available, learning effective representations has attracted attention for improving recommendation performance. Existing approaches \textcolor{black}{often tightly couple multimodal fusion with the architecture and optimization objective of a specific downstream model}, which limits the generality and reusability of the learned representations.
	To address these limitations, we propose LUMIR, a decoupled multimodal representation learning framework based on contrastive learning. Multimodal information collected from heterogeneous data sources is first semantically unified into a shared textual representation space using large language models (LLMs). An interaction-aware fusion mechanism is then designed to model cross-modal interactions, while sequential signals are incorporated as a structural constraint to guide representation learning.
	To evaluate the learned representations, experiments are conducted on a multimodal point-of-interest (POI) dataset under the next-POI recommendation setting across multiple sequential recommendation backbones. The results show that LUMIR consistently outperforms existing multimodal fusion baselines in both accuracy-oriented and beyond-accuracy metrics, demonstrating that decoupling representation learning from downstream optimization provides an effective paradigm for learning reusable multimodal representations. The code is publicly available at \url{https://github.com/wzehui/LUMIR}.
\end{abstract}


\maketitle

\section{Introduction}
With the growing availability of multimodal data, modern recommender systems increasingly leverage cross-modal representation learning in data-driven applications~\cite{MultiRSSurveyLiu2024}. In scenarios such as POI recommendation, item-related information is typically derived from \textcolor{black}{heterogeneous data sources}, including textual descriptions, visual content, and structured attributes. Effectively integrating \textcolor{black}{multimodal tourism-related information from heterogeneous data sources} into a unified and informative representation remains a fundamental challenge~\cite{tsai2019learning}.

Recent advances in large language models (LLMs) provide a promising direction for modeling multimodal tourism-related information by transforming information from heterogeneous data sources into a shared semantic representation space~\cite{an2025llmcentricmultimodalfusionsurvey}. However, \textcolor{black}{heterogeneous data sources may contain semantically diverse, redundant, inconsistent, or only partially informative signals that cannot be fully resolved through semantic unification alone and can introduce noise in multimodal data, thereby degrading the quality of the learned representations.}

Despite these challenges, multimodal representation learning is typically tightly coupled with downstream tasks, where the learned representations are directly optimized based on task-specific performance signals~\cite{Ngiam2011MultimodalDL}. While this design can improve task-specific performance, it introduces several limitations. First, the learned representations become \textcolor{black}{tied to a particular downstream architecture and its optimization process}, limiting their generality and transferability across different tasks or domains~\cite{RLSurveyBengio}. 
Second, \textcolor{black}{joint optimization of multimodal fusion and downstream recommendation models requires representation learning to be repeated when the downstream architecture changes, increasing} computational cost and reducing scalability~\cite{xie2022decoupledinformationfusionsequential}.
More importantly, \textcolor{black}{supervision from downstream objectives} biases multimodal fusion toward task-optimal solutions rather than intrinsic cross-modal relationships, limiting cross-modal interaction modeling~\cite{tsai2019learning}.
\textcolor{black}{These challenges become particularly evident in multimodal POI recommendation, where multimodal tourism-related information collected from heterogeneous data sources must be integrated with sequential user interactions. Such heterogeneous information characteristics motivate the need for a more flexible and decoupled representation learning framework.}

As an important task within POI recommendation, next-POI recommendation aims to predict the next location users will visit based on their historical interaction sequence~\cite{yu2024gnn_poi_survey}. 
Compared to general POI recommendation, next-POI recommendation focuses on modeling short-term sequential patterns and temporal dynamics~\cite{wang2025survey}.
In this setting, multimodal POI information provides complementary signals for capturing contextual characteristics and user preferences~\cite{MMPOIXu2024}, and its effective integration in sequential recommendation has attracted increasing attention in recent work~\cite{EmpowerHuang2022}.

To address these limitations, we propose \textbf{L}LM-based \textbf{U}nified \textbf{M}ultimodal \textbf{I}nformation \textbf{R}epresentation (LUMIR), a decoupled multimodal representation learning framework. \textcolor{black}{Unlike existing approaches that tightly couple representation learning with downstream models, LUMIR learns representations independently of downstream model architectures and their optimization objectives.}
\textcolor{black}{Leveraging the prior knowledge and representation capabilities of LLMs, LUMIR learns unified representations from multimodal tourism-related information collected from heterogeneous data sources (e.g., images, reviews, and metadata). On top of this unified representation space, LUMIR models cross-modal interactions while reducing the influence of redundant information and incorporating sequential signals as structural guidance for representation learning.}
\textcolor{black}{Through stochastic modality augmentation and information compression, the framework is designed to reduce its dependence on noisy or partially informative modality-specific signals without requiring explicit data-quality annotations.}
The learned representations are evaluated on POI recommendation as a downstream task to assess their effectiveness in addressing key challenges in multimodal recommender systems.
To the best of our knowledge, this work is the first to investigate \textcolor{black}{a multimodal representation learning framework decoupled from downstream architectures and their optimization } in the context of next-POI recommendation. The key contributions are summarized as follows:

\begin{itemize}
	
	\item We propose LUMIR, a decoupled multimodal representation learning framework that separates representation learning from \textcolor{black}{the architecture and optimization of} downstream recommendation tasks. 
	This design facilitates a modular approach to multimodal modeling.
	
	\item We propose an LLM-based representation unification framework \textcolor{black}{for multimodal tourism-related information collected from heterogeneous data sources,} in which the encoder is trainable and jointly optimized with the multimodal fusion process, enabling adaptive unified representations tailored to cross-modal interactions.
	
	\item We interpret multimodal representation learning under the information bottleneck framework and propose a unified objective function that jointly integrates modality compression and sequence-aware alignment, enabling the learned representations to capture both cross-modal interactions and sequential structure.
	
	\item We conduct experiments on a multimodal POI dataset under the next-POI recommendation setting, showing that the representations improve performance compared to multimodal baselines.
	
\end{itemize}

\section{Related Work}
This section reviews recent work on task-dependent multimodal representation learning, LLM-based representation unification, and next-POI recommendation.

\paragraph{Multimodal Representation Learning.}
Multimodal representation learning has been extensively studied to learn unified representations from \textcolor{black}{heterogeneous data sources}, where information is distributed across textual, visual, and structured modalities~\cite{Ngiam2011MultimodalDL}. Recent advances therefore move beyond simple alignment to model cross-modal interactions by capturing redundant, unique, and synergistic information across modalities~\cite{WhatDufumier}.

Despite differences in application domains and model architectures, these approaches generally learn multimodal representations \textcolor{black}{through joint optimization with downstream models}, where the fusion process is tightly coupled with downstream objectives. For instance, prior work has evaluated multimodal representations directly based on recommendation performance~\cite{DoRecPomo2025}, learned joint vision-language representations through task-driven supervision~\cite{radford2021learning}, or optimized multimodal models for downstream tasks such as classification and question answering~\cite{alayrac2022flamingo}. 
\textcolor{black}{Such designs produce representations that are tightly optimized together with a particular downstream architecture, limiting their modular reuse across different recommendation models.}

\paragraph{LLM-based Representation Unification.}
Recent advances in LLMs provide a promising direction for multimodal representation learning by enabling multimodal information \textcolor{black}{collected from heterogeneous data sources to be represented within a shared semantic embedding space~\cite{MLLMSurveyLiang2024}.} Such semantic unification helps reduce modality heterogeneity while providing a common foundation for subsequent multimodal representation learning.

Recent studies leverage LLMs as a semantic bridge across application domains. For example, LLM-based semantic enhancement frameworks have been applied to unify textual and visual signals for POI representation learning in the recommendation domain~\cite{wang2025beyond}. In vision-language tasks, models such as BLIP-2 connect visual encoders with LLMs, enabling images to be represented and processed within a language-centric semantic space~\cite{BLIPLi2023}. Similarly, models such as LLaVA align visual and language features via instruction tuning, enabling unified multimodal understanding~\cite{LLAVALiu2023}. 
Building on this prior work, our work uses LLMs to \textcolor{black}{semantically unify multimodal tourism-related information collected from heterogeneous data sources before downstream representation learning.}

\paragraph{Next-POI Recommendation.}
Next-POI recommendation is a key task in tourism recommendation, which aims to predict the next location a user will visit based on historical trajectory~\cite{wang2025survey}. 
POI data is inherently multimodal, with rich information sources associated with each location. Recent work has explored incorporating additional contextual and multimodal information, such as geographical features~\cite{DisenPOIQin2023}, semantic information~\cite{NormalizedAlenezi2022}, and visual content~\cite{Multi-LevelChen2023}. 

With the rapid development of LLMs, their capabilities have increasingly been incorporated into POI recommendation to improve recommendation performance. For example, LLMs have been leveraged to enrich multimodal POI representations by transforming textual and visual information into structured representations~\cite{wang2025beyond}. In addition, methods such as SeCor align LLM-generated representations with collaborative signals~\cite{SeCorWang2024}. 
\textcolor{black}{These works highlight the potential of LLM-based representations for integrating multimodal information, motivating our use of LLM-based encoders within a decoupled representation learning framework.}

\textcolor{black}{Despite these advances, existing approaches generally couple multimodal representation learning with downstream recommendation objectives, making the learned representations \textcolor{black}{dependent on a particular model configuration.} Such \textcolor{black}{end-to-end optimization} may dominate the fusion process and limit the modeling of intrinsic relationships among heterogeneous information sources. Our work therefore investigates a decoupled framework that learns multimodal representations independently of the downstream recommendation objective and evaluates their effectiveness in next-POI recommendation.}

\section{Problem Formulation}\label{Problem Formulation}
This section defines the problem setting and notation for multimodal representation learning and next-POI recommendation.

\subsection{Multimodal Representation Learning}

Let $\mathcal{V} = \{v_1, \ldots, v_{|\mathcal{V}|}\}$ denote a set of recommendable items. 
Each item $v \in \mathcal{V}$ is associated with multimodal information:
\begin{equation}
	\mathbf{m}_v = \{ m_v^{(1)}, \ldots, m_v^{(M)} \},
	\label{eq:multimodal_input}
\end{equation}
where $m_v^{(k)}$ denotes the $k$-th modality, each corresponding to heterogeneous data sources (e.g., text, images, structured data).

The multimodal representation learning problem aims to map heterogeneous modalities into a unified latent space~\cite{radford2021learning,MultiBal2019} and can be abstracted as:
\begin{equation}
	\mathbf{z}_v =  \mathcal{F}(\{m^{(k)}\}_{k=1}^M),
\end{equation}
where $\mathcal{F}(\cdot)$ denotes a multimodal representation function and $\mathbf{z}_v$ is the representation of item $v$.

\begin{figure*}[t]
	\centering
	\includegraphics[width=1\textwidth]{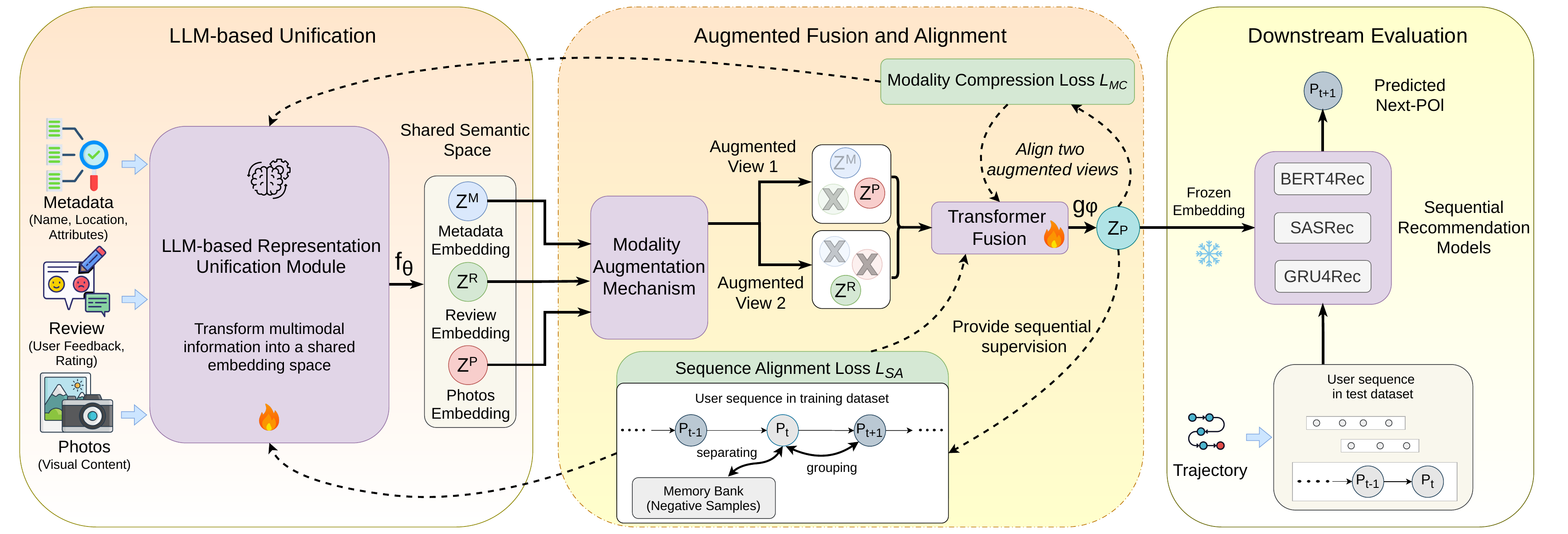}
	\caption{Overview of the proposed LUMIR framework.}
	\label{fig:framework}
\end{figure*}

\subsection{Next-POI Recommendation}

In this work, next-POI recommendation is considered as the downstream application, where each item $v \in \mathcal{V}$ corresponds to a POI.
Let $\mathcal{U} = \{u_1, \ldots, u_{|\mathcal{U}|}\}$ denote the set of users and $\mathcal{P} = \{p_1, \ldots, p_{|\mathcal{P}|}\}$ denote the set of POIs.
For each user $u \in \mathcal{U}$, the historical interaction sequence is defined as:
\begin{equation}
	S_u = \langle p_1, p_2, \ldots, p_T \rangle,
\end{equation}
where $p_t \in \mathcal{P}$ is the POI visited at time step $t$.

Given the observed sequence $S_u$, the next-POI recommendation task aims to predict the next POI $p_{T+1}$.
The prediction is performed based on the sequence $S_u$ and the multimodal representations $\mathbf{z}_p$.

\section{Methodology}
This section presents the proposed LUMIR framework, which consists of three main components: (1) LLM-based representation unification, (2) augmented multimodal representation learning, and (3) downstream evaluation via next-POI recommendation. An overview of the framework architecture is illustrated in Figure~\ref{fig:framework}.

\subsection{LLM-based Representation Unification}



In the multimodal POI dataset used in this work, each POI $p \in \mathcal{P}$ is associated with multiple feature sources, including textual, visual, and structured information. To focus on multimodal representation learning, these heterogeneous feature sources are grouped into three modalities: (1) \textbf{Metadata ($M$):} basic business descriptors such as name and category, structured attributes, and spatial information including geolocation and neighborhood representations based on H3 spatial indexing~\cite{H3_1,H3_2}; (2) \textbf{Review ($R$):} LLM-generated summaries of given user reviews and rating statistics; (3) \textbf{Photo ($P$):} LLM-generated visual content summaries and additional visual features extracted from images.

Based on Eq.~\ref{eq:multimodal_input}, the multimodal information of each POI can be written as:
\begin{equation}
	\mathbf{m}_p = \{ m_p^{(k)} \mid k \in \{M, R, P\} \},
\end{equation}
where $m_p^{(M)}$, $m_p^{(R)}$, and $m_p^{(P)}$ correspond to metadata, review, and photo modalities, respectively.

Following~\cite{wang2025beyond}, each modality is transformed into a structured textual representation and encoded by a tunable LLM-based text encoder to generate modality-specific embeddings:
\begin{equation}
	\mathbf{z}_p^{(k)} = f_{\theta}\big(m_p^{(k)}\big), \quad k \in \{M, R, P\},
\end{equation}
where $f_{\theta}(\cdot)$ denotes the encoding function.

\textcolor{black}{This semantic unification transforms information from different modalities into a shared textual representation space, thereby reducing the heterogeneity of the original modality formats.} These modality-specific embeddings $\{ \mathbf{z}_p^{(k)} \}$ are subsequently used as inputs for multimodal representation learning, where a unified POI representation $\mathbf{z}_p$ is learned by fusion  across modalities.

\subsection{Multimodal Representation Learning}
Given the modality-specific embeddings $\{ \mathbf{z}_p^{(k)} \mid k \in \{M,R,P\} \}$, the goal of LUMIR is to learn a unified POI representation $\mathbf{z}_p$ by integrating information across modalities.
\textcolor{black}{Instead of optimizing multimodal representations through the loss of a downstream model,} LUMIR adopts a self-supervised learning paradigm to capture intrinsic relationships across modalities. In particular, contrastive learning effectively models correlations among multimodal representations by grouping related instances and separating unrelated ones~\cite{MultimodalZhang2024}, making it well suited for decoupled representation learning. Based on this, a contrastive objective is introduced to guide the fusion of modality-specific embeddings.
\paragraph{Modality Compression Loss}

While contrastive learning can align representations across modalities, effective integration of multimodal information goes beyond alignment. In this work, a modality fusion framework proposed in~\cite{WhatDufumier} is adopted. 
In particular, a modality augmentation transformation, denoted by $\mathcal{T}$, is applied to modality-specific embeddings $\{ \mathbf{z}_p^{(k)} \mid k \in \{M,R,P\} \}$, where each embedding corresponding to a modality $k \in \{M,R,P\}$ is randomly perturbed or masked to generate stochastic partial views of the same POI. This encourages the model to learn robust representations by leveraging complementary information across modalities rather than relying on any single modality.
As a result, two independent modality augmentation transformations, denoted by $\mathcal{T}'$ and $\mathcal{T}''$, are applied to $\{ \mathbf{z}_p^{(k)} \mid k \in \{M,R,P\} \}$.
These are then fused via an attention-based multimodal fusion function $g_{\phi}(\cdot)$ to obtain two stochastic views:
\begin{equation}
	\tilde{\mathbf{z}}_p' = g_{\phi}\big(\mathcal{T}'(\mathbf{z}_p^{(k)})\big), \qquad
	\tilde{\mathbf{z}}_p'' = g_{\phi}\big(\mathcal{T}''(\mathbf{z}_p^{(k)})\big).
\end{equation}

These two fused views correspond to different incomplete observations of the same POI and serve as positive pairs in contrastive learning. The modality compression objective enforces consistency between these two fused views while simultaneously aligning each modality-specific embedding with the fused representation. 

Compared with existing approaches that restrict negative sampling within a mini-batch, a practical challenge in this work arises from the high dimensionality of multimodal representations. This leads to increased memory consumption and limits the feasible batch size during training. As contrastive learning relies on a large number of negative samples, small batch sizes reduce the diversity of negatives and can degrade representation quality~\cite{he2020momentum}.
To address this issue, LUMIR employs a memory bank $\mathcal{M}$ to store representations from previous mini-batches and reuse them as additional negative samples~\cite{he2020momentum}. This mechanism effectively enlarges the pool of negative samples without increasing the batch size.
With the Information Noise-Contrastive Estimation (InfoNCE) objective denoted as $\hat{I}_{\mathrm{NCE}}(\cdot,\cdot)$, the modality compression loss is defined as:

{\small
	\begin{equation}
		\mathcal{L}_{\mathrm{MC}}
		=
		- \frac{1}{|\mathcal{M}|}\sum_{p\in\mathcal{M}}
		\left[
		\hat{I}_{\mathrm{NCE}}\!\left(\tilde{\mathbf{z}}_p', \tilde{\mathbf{z}}_p''\right)
		+
		\frac{1}{|\mathcal{K}|}\sum_{k\in\mathcal{K}}
		\bar{I}_{\mathrm{NCE}}\!\left(\mathbf{z}_p^{(k)}, \tilde{\mathbf{z}}_p\right)
		\right],
		\label{eq:MC}
	\end{equation}
}
where 
\begin{equation}
	\bar{I}_{\mathrm{NCE}}\!\left(\mathbf{z}_p^{(k)}, \tilde{\mathbf{z}}_p\right)
	=
	\frac{1}{2}
	\left(
	\hat{I}_{\mathrm{NCE}}\!\left(\mathbf{z}_p^{(k)}, \tilde{\mathbf{z}}_p'\right)
	+
	\hat{I}_{\mathrm{NCE}}\!\left(\mathbf{z}_p^{(k)}, \tilde{\mathbf{z}}_p''\right)
	\right).
\end{equation}


\paragraph{Sequence Alignment Loss}

Inspired by recent advances in applying contrastive learning to sequential recommendation~\cite{UniSRecHou, DuoRecQiu}, a sequence-aware alignment objective is introduced to enhance the multimodal representations. The key assumption is that items appearing within the same user sequence exhibit semantic relatedness. Aligning their representations therefore helps $\mathbf{z}_p$ capture sequential patterns that are useful for sequential modeling. \textcolor{black}{This sequence-derived signal introduces task-relevant structural information into representation learning, while remaining independent of the parameters and optimization objective of any particular downstream sequential recommendation architecture.}

To this end, a sequence-aware contrastive objective is constructed for the next-POI recommendation setting. For a given current POI $p$, positive samples are defined as POIs that appear as subsequent transitions in user sequences, while negative samples are drawn from other POIs stored in a memory bank $\mathcal{M}$. The sequence alignment loss is defined as:

\begin{equation}
	\mathcal{L}_{\mathrm{SA}}
	=
	- \frac{1}{|\mathcal{M}|}\sum_{p\in\mathcal{M}}
	\hat{I}_{\mathrm{NCE}}\!\left(
	\mathbf{z}_p,\,
	\mathcal{P}(p);\,
	\mathcal{M}
	\right),
	\label{eq:SA}
\end{equation}
where $\mathcal{P}(p)$ denotes the set of positive samples, i.e., POIs observed as next-step transitions for the current POI $p$ across user sequences, and negative samples are sampled from $\mathcal{M}$.
\paragraph{Joint Objective}
The two contrastive objectives introduced above provide complementary learning signals for multimodal POI representation learning. To better understand how they jointly shape the learned representation, the framework can be interpreted from an information-theoretic perspective. Let $X_p$ denote the original multimodal input of POI $p$, defined as the collection of modality-specific information, i.e., $X_p = \{ m_p^{(k)} \mid k \in \{M, R, P\} \}$. Multimodal representation learning can be viewed as optimizing the mutual information between $X_p$ and $\mathbf{z}_p$, aiming to learn a compact representation that suppresses redundant information while retaining task-relevant signals. Following the Information Bottleneck (IB) principle~\cite{DVIBZhao, AttentionNagrani, MultimodalMai}, the objective can be conceptually formulated in terms of mutual information $I$ as:

\begin{equation}
	\mathcal{L}_{\mathrm{IB}}
	= I(X_p;\mathbf{z}_p) - \beta I(\mathbf{z}_p; y),
	\label{eq:ib_objective}
\end{equation}
where $y$ denotes the prediction target in the next-POI recommendation task.

In this work, the prediction target $y$ cannot be directly used during representation learning due to the decoupled training paradigm. 
To address this issue, the relational structure within user sequences is used as a surrogate signal. The underlying intuition is that the prediction target $y$ is typically derived from, or conditioned on, the user sequence. Therefore, representations that capture sequence-level dependencies are likely to retain information relevant to the downstream task.
This relationship can be interpreted through the Data Processing Inequality (DPI), which states that processing a signal cannot increase the information it contains about an underlying target. Under this assumption, maximizing the mutual information between the representation $\mathbf{z}_p$ and sequence-level signals can serve as a surrogate objective for preserving task-relevant information.
While this surrogate is not strictly equivalent to directly maximizing $I(\mathbf{z}_p; y)$, it provides a tractable and informative optimization signal under the decoupled setting.

Based on Eq.~\ref{eq:ib_objective}, LUMIR approximates the IB objective using contrastive learning. 
Specifically, the modality compression term is formulated as Eq.~\ref{eq:MC}. By constructing multiple augmented views of the same POI and enforcing consistency between fused representations and modality-specific embeddings, $\mathcal{L}_{\mathrm{MC}}$ encourages the model to retain shared and complementary information across modalities while suppressing redundant signals.
Similarly, the sequence alignment term is implemented as Eq.~\ref{eq:SA}, which provides a tractable lower-bound estimator of mutual information between paired variables~\cite{he2020momentum}. As a result, minimizing $\mathcal{L}_{\mathrm{SA}}$ implicitly maximizes the mutual information between $\mathbf{z}_p$ and the sequence-derived target signal.
The final training objective is formulated as:
\begin{equation}
	\mathcal{L}_{\mathrm{Total}}
	= \lambda_{\mathrm{MC}} \cdot \mathcal{L}_{\mathrm{MC}}
	+ \lambda_{\mathrm{SA}} \cdot \mathcal{L}_{\mathrm{SA}},
\end{equation}
where $\lambda_{\mathrm{MC}}$ and $\lambda_{\mathrm{SA}}$ balance multimodal compression and sequence-aware alignment, respectively.

Differently from existing studies \cite{IISANFu, MISSRecWang} that couple multimodal fusion with downstream sequential recommendation models via joint optimization, LUMIR adopts a decoupled representation learning paradigm. The representation is learned by jointly optimizing the multimodal compression loss $\mathcal{L}_{\mathrm{MC}}$ and the sequence-aware alignment loss $\mathcal{L}_{\mathrm{SA}}$. The optimization process updates the parameters of the tunable LLM-based encoder $f_{\theta}(\cdot)$ and the multimodal fusion function $g_{\phi}(\cdot)$, which together define the mapping from modality-specific inputs to the unified representation.
The final embedding configuration is selected based on a proxy objective defined in this work, which measures how frequently similar embeddings co-occur within the same user sequence, reflecting the ability of the learned representations to capture sequence-level consistency.

\subsection{Downstream Sequential Recommendation}
After modality fusion, the learned POI representations $\tilde{\mathbf{z}}_p$ are evaluated as input representations for downstream next-POI recommendation. In this work, BERT4Rec~\cite{sun2019bert4rec}, SASRec~\cite{kang2018self}, and GRU4Rec~\cite{hidasi2015session} are employed as representative sequential recommendation models. To incorporate the learned POI embeddings, the original ID-based embedding layer in these models is replaced with a pre-initialized embedding matrix constructed from $\tilde{\mathbf{z}}_p$, following prior work~\cite{ImprovingSequentialBoz2025}.

Given a user interaction sequence $s = [p_1, p_2, \dots, p_t]$, each POI $p_i$ is represented by its embedding $\tilde{\mathbf{z}}_{p_i}$. The sequential model takes the embedding sequence as input and learns to predict the next POI $p_{t+1}$ based on the history. During this stage, the POI embeddings $\tilde{\mathbf{z}}_p$ remain fixed, and the sequential model is trained to capture temporal or sequential dependencies over the embedding sequence.

\section{Experiments}
This section describes the experimental setup of the LUMIR framework, including the dataset, implementation details and baseline methods. 
Multimodal POI representations are learned using different fusion strategies and evaluated on downstream next-POI recommendation under a unified setting.

\subsection{Dataset}
Experiments are conducted on the Yelp multimodal dataset\footnote{\url{https://huggingface.co/datasets/wzehui/Yelp-Multimodal-Recommendation}}, which is one of the publicly available multimodal datasets in the tourism domain accessible during this study. It contains rich business-related information, including metadata, user reviews, visual content, and user interaction data organized into sequential visit trajectories.\footnote{In addition to the Yelp dataset, we also conducted experiments on a multimodal Instagram dataset constructed in this work. Due to its user-centric nature and differing data characteristics, the experimental setting differs from the business-centric scenario considered in this work. A detailed investigation of this setting is left for future work.}
The statistics of the preprocessed dataset are summarized in Table~\ref{tab:dataset}.

\begin{table}[h]
	
	\centering
	\scriptsize
	\setlength{\tabcolsep}{3pt}
	\renewcommand{\arraystretch}{0.95}
	
	\caption{Statistics of the Yelp multimodal POI dataset}
	\label{tab:dataset}
	
	\begin{tabular*}{\linewidth}{@{\extracolsep{\fill}} l c l c}
		\toprule
		\textbf{Metric} & \textbf{Value} & \textbf{Metric} & \textbf{Value} \\
		\midrule
		\#Interactions & 509,892 & Avg. User Seq. Length & 23.0 \\
		\#Businesses & 16,702 & Avg. Reviews per Business & 30.5 \\
		\#Users & 22,207 & Avg. Photos per Business & 9.0 \\
		Sparsity & 99.86\% & Businesses with Photos & 80\% \\
		\bottomrule
	\end{tabular*}
	
\end{table}

\subsection{Implementation Details}
The experimental pipeline follows a decoupled setting. First, all modality-specific information is transformed into a unified embedding space via LLM-based representation unification. In this work, \texttt{Linq-Embed-Mistral}\footnote{\url{https://huggingface.co/Linq-AI-Research/Linq-Embed-Mistral}} is employed as a tunable LLM encoder to obtain modality-specific embeddings. The model is selected based on its strong embedding performance on the MTEB benchmark.
POI embeddings are then learned independently for each method. 
During this stage, hyperparameters for representation learning are tuned via grid search, where the best-performing embeddings are selected based on the proxy objective. 
\textcolor{black}{For LUMIR, modality augmentation is implemented through modality-level hard masking. During the first three warm-up epochs, all three modalities are retained. Afterwards, the probability of retaining the full modality set decreases linearly during training, with a lower bound of 0.2. Otherwise, the number of modalities to be retained is randomly selected with equal probability. The two augmented views are generated independently using this procedure.}

The learned embeddings are subsequently fixed and used as input to downstream sequential recommendation models. 
\textcolor{black}{All downstream models are optimized using the Adam optimizer with a cross-entropy objective over POIs. BERT4Rec is trained through masked-item prediction, whereas SASRec and GRU4Rec are trained for next-item prediction.}
For the downstream models, hyperparameter tuning is performed using Optuna with 50 trials, where the best configuration is selected based on validation performance measured by NDCG@20. 
After identifying the best-performing hyperparameter configuration for each method, both LUMIR and the baselines are retrained under their respective optimal configurations for 10 runs with different random initializations. The final performance is reported as the average over these runs after removing the highest and lowest values to mitigate the effect of outliers.

All experiments are conducted on NVIDIA A40 (48GB) GPU for the representation learning stage and NVIDIA A800 (80GB) GPU for the downstream next-POI recommendation tasks.

\subsection{Baselines}\label{Baselines}
In this work, by applying the POI embeddings generated from various baselines  to the same downstream recommendation task, LUMIR can be systematically evaluated under a unified experimental setting. \textcolor{black}{Although recent approaches such as SeCor similarly leverage LLMs to integrate different information sources, SeCor additionally incorporates collaborative information derived from user interactions, which prevents a fair comparison with LUMIR under the same input conditions and is therefore not included as a baseline.} In this work, four strategies based on different principles are adopted to construct POI representations:

\paragraph{Trajectory-Only Strategy.}
This strategy represents an ID-based baseline that does not incorporate any additional POI representations derived from multimodal information. In this setting, the next-POI recommendation task relies on the sequential patterns observed in users’ historical visit trajectories.

\paragraph{Full Concatenation Strategy.} 
This strategy represents a parameter-free feature-level fusion strategy that concatenates embeddings from different modalities while preserving the complete information of each modality \cite{wang2025beyond}. In this study, it serves as a baseline to evaluate the performance that can be achieved through feature-level fusion without explicitly modeling interactions between modalities.

\paragraph{Modality-Weighted Strategy.}
This strategy extends the Full Concatenation Strategy by adjusting different modality weights.
The final POI representation is obtained through a weighted concatenation of modality embeddings. This strategy maintains a simple fusion structure while introducing explicit modeling of modality importance, which is commonly studied in multimodal fusion methods \cite{MultiBal2019}.

\paragraph{FLAVA-based Strategy.}
FLAVA is a vision–language model that learns joint representations via a shared encoder and cross-modal alignment objectives \cite{Singh_2022_CVPR}.
In the dataset used in this work, modality-specific information (e.g., reviews and images) is summarized into textual representations by LLMs. The model is adapted to take the textual representations of the three modalities as input, while preserving the input structure as LUMIR. 
During fusion, attention mechanisms capture interactions across modalities within a unified encoding space, enabling implicit cross-modal integration beyond simple concatenation.

\paragraph{CoMM-based Strategy.}
The CoMM strategy learns multimodal representations by aligning modality-specific embeddings in a shared space \cite{WhatDufumier}. 
It focuses on modality-level alignment, which is conceptually similar to LUMIR, but does not incorporate sequential semantics or address the limitation of insufficient negative samples under small batch sizes.

\section{Results}
This section evaluates LUMIR framework from two perspectives. (1) The multimodal fusion process in LUMIR is analyzed to understand how meaningful structures emerge during representation learning. (2) The learned embeddings are assessed through their performance on downstream next-POI recommendation tasks.

\subsection{Multimodal Fusion Dynamics}
Due to the absence of direct supervision from downstream recommendation tasks, the quality of the learned representations in the multimodal fusion stage emerges from the training dynamics of the representation learning process itself. 
Different from existing unsupervised embedding methods that rely on fixed iterative propagation~\cite{rychalska2021cleora}, LUMIR involves learnable multimodal fusion with complex training dynamics, making the training process a critical factor in determining representation quality.
As a result, insufficient training may lead to under-expressive representations that fail to capture meaningful multimodal structures, while excessive training may overfit dominant interaction patterns, biasing representations toward popular POIs and under-representing long-tail items~\cite{WU2024121429}. 
Therefore, two aspects are examined: (a) how the learning process improves the structural properties of the embeddings; and (b) how these changes affect downstream recommendation performance.
Representations obtained from direct concatenation of untrained modality features are used as an initial status and compared with those learned at different epochs (5–25). 
The range of epochs is selected to capture the evolution of representations at different training stages within a practical computational budget.

\begin{figure}[!tbh]
	\centering
	
	\begin{subfigure}{0.3\linewidth}
		\includegraphics[width=\linewidth]{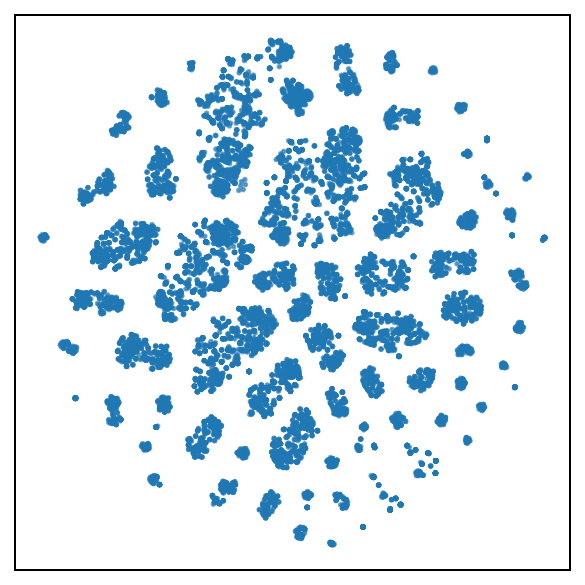}
		\caption{Concatenation}
	\end{subfigure}
	\hfill
	\begin{subfigure}{0.3\linewidth}
		\includegraphics[width=\linewidth]{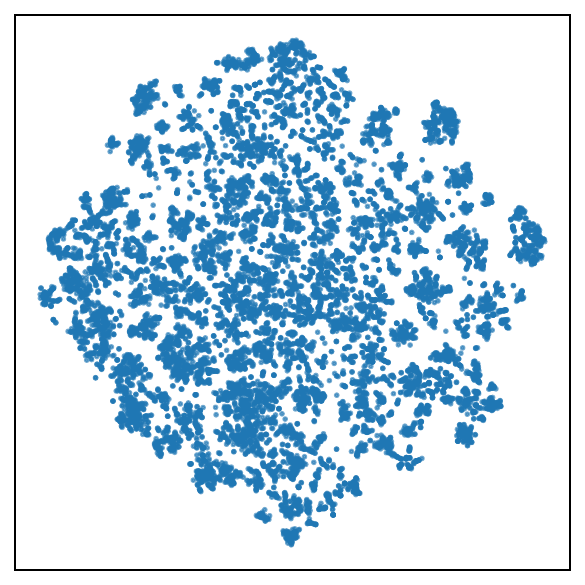}
		\caption{Epoch 10}
	\end{subfigure}
	\hfill
	\begin{subfigure}{0.3\linewidth}
		\includegraphics[width=\linewidth]{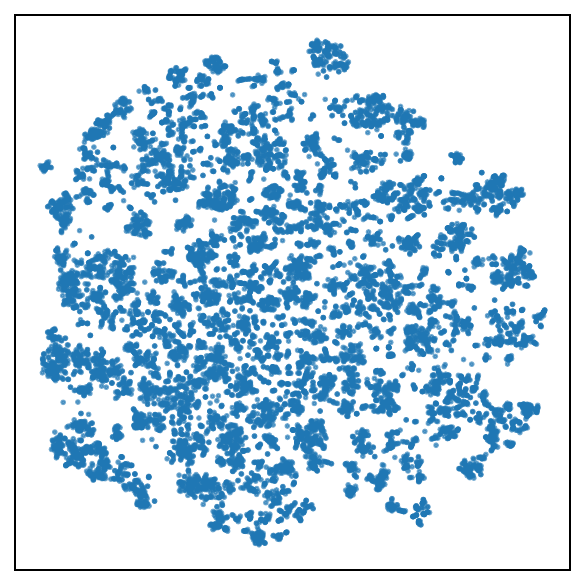}
		\caption{Epoch 25}
	\end{subfigure}
	
	\caption{Visualization of Embeddings Across Training.}
	\label{fig:tsne_dynamics}
	
\end{figure}

\paragraph{Embedding Structure Evolution.} As shown in Figure~\ref{fig:tsne_dynamics}, the learned embeddings form more coherent clusters, as visualized using t-SNE. As representation learning progresses, the embedding space evolves from fragmented, attribute-driven groupings to more coherent and semantically structured clusters. While these visualizations provide important insights into the structural evolution of the embedding space, quantitative evaluation is still required to fully assess representation quality.

\begin{figure}[!tbh]
	\centering
	
	\begin{subfigure}[t]{0.43\linewidth}
		\centering
		\includegraphics[width=\linewidth]{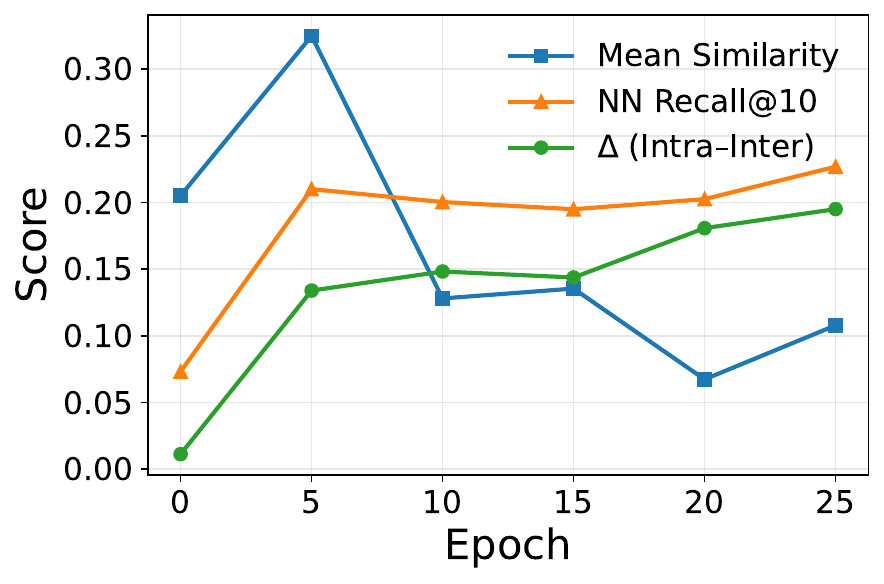}
		\caption{Structural properties of learned representations}
		\label{fig:embedding_dynamics}
	\end{subfigure}
	\hfill
	\begin{subfigure}[t]{0.43\linewidth}
		\centering
		\includegraphics[width=\linewidth]{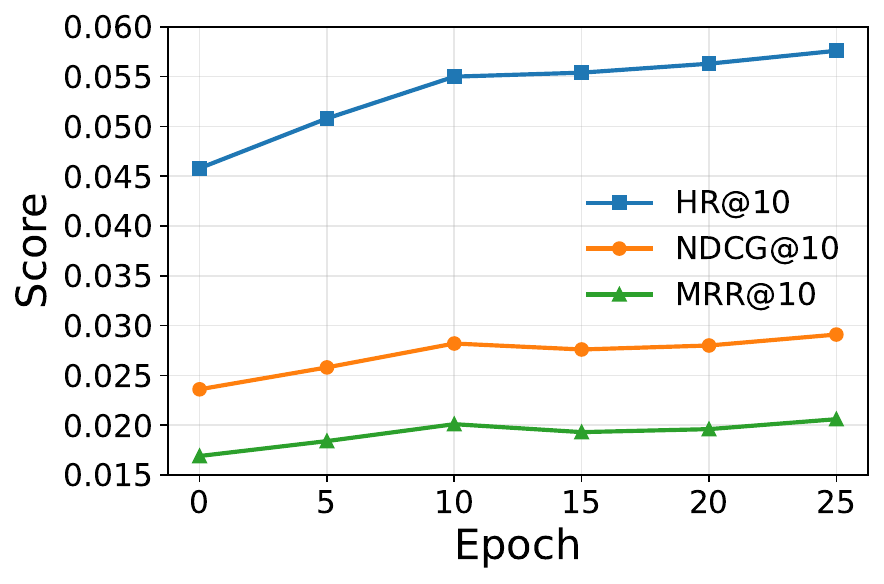}
		\caption{Downstream performance across epochs}
		\label{fig:downstream_dynamics}
	\end{subfigure}
	
	\caption{Evolution of representation structure and downstream performance during representation learning}
	\label{fig:training_dynamics}
\end{figure}

To this end, three metrics are used to characterize the structural properties of the learned embeddings: (1) Mean Similarity, which reflects the overall compactness of the embedding space; (2) Nearest Neighbor Recall@10 (NN Recall@10), which measures local neighborhood consistency~\cite{recall}; and (3) the Intra–Inter Similarity Gap ($\Delta$), which captures the separation between intra- and inter-session similarities~\cite{delta}.
As shown in Figure~\ref{fig:embedding_dynamics}, Mean Similarity increases at early epochs and then decreases, indicating a shift from overly compact representations to a more dispersed structure. In contrast, NN Recall@10 and $\Delta$ exhibit upward trends, suggesting improved local consistency and stronger discriminability between intra- and inter-session relationships as representation learning progresses.

\paragraph{Impact on Downstream Performance.} While the above analysis provides insights into the structural properties of the learned embeddings, improvements in embedding structure do not necessarily translate into gains in downstream performance. Therefore, this section focuses on the evolution of downstream performance during representation training, rather than final evaluation results. As shown in Figure~\ref{fig:downstream_dynamics}, all metrics exhibit consistent improvements during early training stages. HR@10, NDCG@10, and MRR@10 increase notably from epoch 0 to epoch 10, indicating that representation learning effectively enhances recommendation performance. After this stage, performance gains become marginal, with slight fluctuations observed across later epochs.

Therefore, considering the trade-off between training cost, embedding structure, and downstream performance, epoch 15 is selected as the final configuration. At this point, the learned representations achieve a balanced state, where intra–inter separation is well established while downstream performance remains stable.

\begin{table*}[!tbp]
	\centering
	\tiny
	\setlength{\tabcolsep}{4pt}
	\renewcommand{\arraystretch}{0.8}
	
	\caption{Performance on the Yelp dataset with different sequential recommendation backbones}
	\label{tab:overall_performance}
	\begin{subtable}{\textwidth}
		
		\centering
		
		\caption{BERT4Rec Backbone}
		
		\begin{tabular*}{\textwidth}{@{\extracolsep{\fill}}
				>{\centering\arraybackslash}m{0.8cm}   
				>{\raggedright\arraybackslash}p{2cm}
				>{\centering\arraybackslash}m{1.85cm}
				>{\centering\arraybackslash}m{1.85cm}
				>{\centering\arraybackslash}m{1.85cm}
				>{\centering\arraybackslash}m{1.85cm}
				>{\centering\arraybackslash}m{1.85cm}
				>{\centering\arraybackslash}m{1.85cm}
			}
			\toprule
			\multirow{9}{*}{
				\parbox[c][5cm][c]{0.8cm}{
					\centering
					\rotatebox[origin=c]{90}{\textit{Accuracy-oriented}}
				}
			}
			&\multirow{2}{*}{\parbox[c]{2cm}{\centering Embedding}}
			& \multicolumn{3}{c}{\textit{@10}}
			& \multicolumn{3}{c}{\textit{@20}} \\
			\cmidrule(lr){3-5} \cmidrule(lr){6-8}
			&  
			& NDCG & HitRate & MRR
			& NDCG & HitRate & MRR \\
			\cmidrule(l{2pt}r{0pt}){2-8}
			
			&Trajectory-Only
			& \shortstack{0.01940 $\pm$ 0.0019 \\ (-20.0\%)}
			& \shortstack{0.03901 $\pm$ 0.0051 \\ (-18.6\%)}
			& \shortstack{0.01353 $\pm$ 0.0012 \\ (-21.0\%)}
			& \shortstack{0.02618 $\pm$ 0.0021 \\ (-16.7\%)}
			& \shortstack{0.06608 $\pm$ 0.0058 \\ (-13.6\%)}
			& \shortstack{0.01535 $\pm$ 0.0012 \\ (-19.6\%)} \\ [0.8em]
			
			&Full Concatenation
			& \shortstack{0.02302 $\pm$ 0.0016 \\ (-5.1\%)}
			& \shortstack{0.04399 $\pm$ 0.0042 \\ (-8.2\%)}
			& \shortstack{0.01671 $\pm$ 0.0011 \\ (-2.5\%)}
			& \shortstack{0.02941 $\pm$ 0.0016 \\ (-6.5\%)}
			& \shortstack{0.06961 $\pm$ 0.0051 \\ (-9.0\%)}
			& \shortstack{0.01841 $\pm$ 0.0010 \\ (-3.5\%)} \\ [0.8em]
			
			&Modality-Weighted
			& \shortstack{\underline{0.02426 $\pm$ 0.0018} \\ (+0.0\%)}
			& \shortstack{\underline{0.04794 $\pm$ 0.0039} \\ (+0.0\%)}
			& \shortstack{\underline{0.01713 $\pm$ 0.0015} \\ (+0.0\%)}
			& \shortstack{\underline{0.03144 $\pm$ 0.0022} \\ (+0.0\%)}
			& \shortstack{\underline{0.07647 $\pm$ 0.0066} \\ (+0.0\%)}
			& \shortstack{\underline{0.01908 $\pm$ 0.0015} \\ (+0.0\%)} \\ [0.8em]
			
			&FLAVA-Based
			& \shortstack{0.02263 $\pm$ 0.00086 \\ (-6.7\%)}
			& \shortstack{0.04686 $\pm$ 0.0021 \\ (-2.3\%)}
			& \shortstack{0.01540 $\pm$ 0.0014 \\ (-10.1\%)}
			& \shortstack{0.02875 $\pm$ 0.0012 \\ (-8.6\%)}
			& \shortstack{0.07127 $\pm$ 0.0030 \\ (-6.8\%)}
			& \shortstack{0.01704 $\pm$ 0.0015 \\ (-10.7\%)} \\ [0.8em]
			
			&CoMM-Based
			& \shortstack{0.02341 $\pm$ 0.00064 \\ (-3.5\%)}
			& \shortstack{0.04526 $\pm$ 0.0016 \\ (-5.6\%)}
			& \shortstack{0.01681 $\pm$ 0.0010 \\ (-1.9\%)}
			& \shortstack{0.02993 $\pm$ 0.0012 \\ (-4.8\%)}
			& \shortstack{0.07123 $\pm$ 0.0040 \\ (-6.9\%)}
			& \shortstack{0.01857 $\pm$ 0.0011 \\ (-2.7\%)} \\ [0.8em]
			
			&LUMIR$_{\mathrm{MC}}$
			& \shortstack{0.02663 $\pm$ 0.0014 \\ (+9.8\%)}
			& \shortstack{0.05304 $\pm$ 0.0021 \\ (+10.6\%)}
			& \shortstack{0.01873 $\pm$ 0.0015 \\ (+9.3\%)}
			& \shortstack{0.03591 $\pm$ 0.0017 \\ (+14.2\%)}
			& \shortstack{0.08988 $\pm$ 0.0056 \\ (+17.5\%)}
			& \shortstack{0.02126 $\pm$ 0.0015 \\ (+11.4\%)} \\ [0.8em]
			
			&LUMIR$_{\mathrm{MC}+\mathrm{SA}}$
			& \shortstack{\textbf{0.02915 $\pm$ 0.0029} \\ (+20.2\%)}
			& \shortstack{\textbf{0.05956 $\pm$ 0.0051} \\ (+24.2\%)}
			& \shortstack{\textbf{0.02010 $\pm$ 0.0025} \\ (+17.3\%)}
			& \shortstack{\textbf{0.03896 $\pm$ 0.0029} \\ (+23.9\%)}
			& \shortstack{\textbf{0.09857 $\pm$ 0.0057} \\ (+28.9\%)}
			& \shortstack{\textbf{0.02276 $\pm$ 0.0025} \\ (+19.3\%)} \\
			
			\midrule
		\end{tabular*}
		
	\end{subtable}
	
	\vspace{0.5em}

	\begin{subtable}{\textwidth}
		
		\centering
		
		\caption{SASRec Backbone}
		
		\begin{tabular*}{\textwidth}{@{\extracolsep{\fill}}
				>{\centering\arraybackslash}m{0.8cm}   
				>{\raggedright\arraybackslash}p{2cm}
				>{\centering\arraybackslash}m{1.85cm}
				>{\centering\arraybackslash}m{1.85cm}
				>{\centering\arraybackslash}m{1.85cm}
				>{\centering\arraybackslash}m{1.85cm}
				>{\centering\arraybackslash}m{1.85cm}
				>{\centering\arraybackslash}m{1.85cm}
			}
			\toprule
			\multirow{9}{*}{
				\parbox[c][5cm][c]{0.8cm}{
					\centering
					\rotatebox[origin=c]{90}{\textit{Accuracy-oriented}}
				}
			}
			&\multirow{2}{*}{\parbox[c]{2cm}{\centering Embedding}}
			& \multicolumn{3}{c}{\textit{@10}}
			& \multicolumn{3}{c}{\textit{@20}} \\
			\cmidrule(lr){3-5} \cmidrule(lr){6-8}
			&  
			& NDCG & HitRate & MRR
			& NDCG & HitRate & MRR \\
			\cmidrule(l{2pt}r{0pt}){2-8}
			
			&Trajectory-Only
			& \shortstack{0.01789 $\pm$ 0.0016 \\ (-21.5\%)}
			& \shortstack{0.03304 $\pm$ 0.0043 \\ (-21.5\%)}
			& \shortstack{0.01331 $\pm$ 0.00090 \\ (-21.8\%)}
			& \shortstack{0.02352 $\pm$ 0.0023 \\ (-18.1\%)}
			& \shortstack{0.05530 $\pm$ 0.0068 \\ (-17.9\%)}
			& \shortstack{0.01485 $\pm$ 0.0011 \\ (-20.2\%)} \\ [0.8em]
			
			&Full Concatenation
			& \shortstack{0.01731 $\pm$ 0.0025 \\ (-24.0\%)}
			& \shortstack{0.03358 $\pm$ 0.0049 \\ (-20.2\%)}
			& \shortstack{0.01242 $\pm$ 0.0019 \\ (-27.0\%)}
			& \shortstack{0.02322 $\pm$ 0.0033 \\ (-19.2\%)}
			& \shortstack{0.05720 $\pm$ 0.0087 \\ (-15.1\%)}
			& \shortstack{0.01401 $\pm$ 0.0021 \\ (-24.7\%)} \\[0.8em]
			
			&Modality-Weighted
			& \shortstack{0.02167 $\pm$ 0.0022 \\ (-4.9\%)}
			& \shortstack{0.04200 $\pm$ 0.0032 \\ (-0.2\%)}
			& \shortstack{0.01557 $\pm$ 0.0021 \\ (-8.5\%)}
			& \shortstack{0.02802 $\pm$ 0.0025 \\ (-2.4\%)}
			& \shortstack{\underline{0.06734 $\pm$ 0.0063} \\ (+0.0\%)}
			& \shortstack{0.01729 $\pm$ 0.0021 \\ (-7.1\%)} \\ [0.8em]
			
			&FLAVA-Based
			& \shortstack{\underline{0.02279 $\pm$ 0.00096} \\ (+0.0\%)}
			& \shortstack{\underline{0.04209 $\pm$ 0.0023} \\ (+0.0\%)}
			& \shortstack{\underline{0.01701 $\pm$ 0.0010} \\ (+0.0\%)}
			& \shortstack{\underline{0.02872 $\pm$ 0.0011} \\ (+0.0\%)}
			& \shortstack{0.06571 $\pm$ 0.0044 \\ (-2.4\%)}
			& \shortstack{\underline{0.01861 $\pm$ 0.00089} \\ (+0.0\%)} \\ [0.8em]
			
			&CoMM-Based
			& \shortstack{0.01770 $\pm$ 0.0028 \\ (-22.3\%)}
			& \shortstack{0.03458 $\pm$ 0.0057 \\ (-17.8\%)}
			& \shortstack{0.01262 $\pm$ 0.0020 \\ (-25.8\%)}
			& \shortstack{0.02366 $\pm$ 0.0028 \\ (-17.6\%)}
			& \shortstack{0.05829 $\pm$ 0.0065 \\ (-13.4\%)}
			& \shortstack{0.01424 $\pm$ 0.0020 \\ (-23.5\%)} \\ [0.8em]
			
			&LUMIR$_{\mathrm{MC}}$
			& \shortstack{\textbf{0.02538 $\pm$ 0.0021} \\ (+11.4\%)}
			& \shortstack{\textbf{0.05241 $\pm$ 0.0038} \\ (+24.5\%)}
			& \shortstack{\textbf{0.01733 $\pm$ 0.0017} \\ (+1.9\%)}
			& \shortstack{\textbf{0.03361 $\pm$ 0.0020} \\ (+17.0\%)}
			& \shortstack{\textbf{0.08517 $\pm$ 0.0039} \\ (+26.5\%)}
			& \shortstack{\textbf{0.01955 $\pm$ 0.0017} \\ (+5.1\%)} \\ [0.8em]
			
			&LUMIR$_{\mathrm{MC}+\mathrm{SA}}$
			& \shortstack{0.02317 $\pm$ 0.0025 \\ (+1.7\%)}
			& \shortstack{0.04480 $\pm$ 0.0056 \\ (+6.4\%)}
			& \shortstack{0.01664 $\pm$ 0.0017 \\ (-2.2\%)}
			& \shortstack{0.03032 $\pm$ 0.0031 \\ (+5.6\%)}
			& \shortstack{0.07332 $\pm$ 0.0081 \\ (+8.9\%)}
			& \shortstack{0.01857 $\pm$ 0.0019 \\ (-0.2\%)} \\
			\midrule
		\end{tabular*}

	\end{subtable}
	
	\vspace{0.5em}

	\begin{subtable}{\textwidth}
		
		\centering
		
		\caption{GRU4Rec Backbone}
		
		\begin{tabular*}{\textwidth}{@{\extracolsep{\fill}}
				>{\centering\arraybackslash}m{0.8cm}   
				>{\raggedright\arraybackslash}p{2cm}
				>{\centering\arraybackslash}m{1.85cm}
				>{\centering\arraybackslash}m{1.85cm}
				>{\centering\arraybackslash}m{1.85cm}
				>{\centering\arraybackslash}m{1.85cm}
				>{\centering\arraybackslash}m{1.85cm}
				>{\centering\arraybackslash}m{1.85cm}
			}
			\toprule
			\multirow{9}{*}{
				\parbox[c][5cm][c]{0.8cm}{
					\centering
					\rotatebox[origin=c]{90}{\textit{Accuracy-oriented}}
				}
			}
			&\multirow{2}{*}{\parbox[c]{2cm}{\centering Embedding}}
			& \multicolumn{3}{c}{\textit{@10}}
			& \multicolumn{3}{c}{\textit{@20}} \\
			\cmidrule(lr){3-5} \cmidrule(lr){6-8}
			&  
			& NDCG & HitRate & MRR
			& NDCG & HitRate & MRR \\
			\cmidrule(l{2pt}r{0pt}){2-8}
			
			& \shortstack[c]{Trajectory-Only}
			& \shortstack{0.01650 $\pm$ 0.0016 \\ (-13.5\%)}
			& \shortstack{0.03376 $\pm$ 0.0040 \\ (-10.1\%)}
			& \shortstack{0.01132 $\pm$ 0.0010 \\ (-16.3\%)}
			& \shortstack{0.02159 $\pm$ 0.0016 \\ (-14.5\%)}
			& \shortstack{0.05395 $\pm$ 0.0041 \\ (-14.7\%)}
			& \shortstack{0.01271 $\pm$ 0.0010 \\ (-16.3\%)} \\ [0.8em]
			
			&Full Concatenation
			& \shortstack{0.01464 $\pm$ 0.0010 \\ (-23.2\%)}
			& \shortstack{0.03068 $\pm$ 0.0017 \\ (-18.3\%)}
			& \shortstack{0.009856 $\pm$ 0.0012 \\ (-27.1\%)}
			& \shortstack{0.01906 $\pm$ 0.0013 \\ (-24.5\%)}
			& \shortstack{0.04843 $\pm$ 0.0023 \\ (-23.5\%)}
			& \shortstack{0.01104 $\pm$ 0.0012 \\ (-27.3\%)} \\ [0.8em]
			
			&Modality-Weighted
			& \shortstack{0.01814 $\pm$ 0.0019 \\ (-4.9\%)}
			& \shortstack{0.03621 $\pm$ 0.0049 \\ (-3.6\%)}
			& \shortstack{0.01273 $\pm$ 0.0012 \\ (-5.8\%)}
			& \shortstack{0.02495 $\pm$ 0.0018 \\ (-1.2\%)}
			& \shortstack{\underline{\textbf{0.06327 $\pm$ 0.0046}} \\ (+0.0\%)}
			& \shortstack{0.01459 $\pm$ 0.0011 \\ (-3.9\%)} \\ [0.8em]
			
			&FLAVA-Based
			& \shortstack{\underline{0.01907 $\pm$ 0.0017} \\ (+0.0\%)}
			& \shortstack{\underline{0.03756 $\pm$ 0.0039} \\ (+0.0\%)}
			& \shortstack{\underline{0.01352 $\pm$ 0.0014} \\ (+0.0\%)}
			& \shortstack{\underline{0.02525 $\pm$ 0.0016} \\ (+0.0\%)}
			& \shortstack{0.06218 $\pm$ 0.0042 \\ (-1.7\%)}
			& \shortstack{\underline{0.01519 $\pm$ 0.0014} \\ (+0.0\%)} \\ [0.8em]
			
			&CoMM-Based
			& \shortstack{0.01748 $\pm$ 0.0014 \\ (-8.3\%)}
			& \shortstack{0.03476 $\pm$ 0.0031 \\ (-7.5\%)}
			& \shortstack{0.01230 $\pm$ 0.0016 \\ (-9.0\%)}
			& \shortstack{0.02314 $\pm$ 0.0014 \\ (-8.4\%)}
			& \shortstack{0.05730 $\pm$ 0.0043 \\ (-9.4\%)}
			& \shortstack{0.01384 $\pm$ 0.0015 \\ (-8.9\%)} \\ [0.8em]
			
			&LUMIR$_{\mathrm{MC}}$
			& \shortstack{\textbf{0.02103 $\pm$ 0.0024} \\ (+10.3\%)}
			& \shortstack{\textbf{0.04191 $\pm$ 0.0035} \\ (+11.6\%)}
			& \shortstack{\textbf{0.01476 $\pm$ 0.0023} \\ (+9.2\%)}
			& \shortstack{\textbf{0.02629 $\pm$ 0.0023} \\ (+4.1\%)}
			& \shortstack{0.06309 $\pm$ 0.0033 \\ (-0.3\%)}
			& \shortstack{\textbf{0.01615 $\pm$ 0.0023} \\ (+6.3\%)} \\ [0.8em]
			
			&LUMIR$_{\mathrm{MC+SA}}$
			& \shortstack{0.01758 $\pm$ 0.0022 \\ (-7.8\%)}
			& \shortstack{0.03467 $\pm$ 0.0043 \\ (-7.7\%)}
			& \shortstack{0.01244 $\pm$ 0.0019 \\ (-8.0\%)}
			& \shortstack{0.02235 $\pm$ 0.0024 \\ (-11.5\%)}
			& \shortstack{0.05358 $\pm$ 0.0054 \\ (-15.3\%)}
			& \shortstack{0.01374 $\pm$ 0.0019 \\ (-9.5\%)} \\
			\midrule
		\end{tabular*}
		
	\end{subtable}
	
	{\footnotesize
		\textbf{Bold} and \underline{underline} indicate the highest mean value for each metric and the best-performing baseline, respectively. Results are averaged over 8 runs after removing the highest and lowest from 10 independent runs. Percentages denote relative performance differences with respect to the best-performing baseline.
	}
\end{table*}

\begin{table*}[!t]
	
	\centering
	\tiny
	\setlength{\tabcolsep}{2.5pt}
	\renewcommand{\arraystretch}{1}
	
	\caption{Statistical significance results across backbones}
	\label{tab:significance}
	
	\begin{tabular*}{\textwidth}{
			@{\extracolsep{\fill}}
			l
			lll @{\hspace{20pt}}
			lll @{\hspace{20pt}}
			lll
		}
		\toprule
		
		& \multicolumn{3}{c}{\textbf{BERT4Rec}} 
		& \multicolumn{3}{c}{\textbf{SASRec}} 
		& \multicolumn{3}{c}{\textbf{GRU4Rec}} \\
		
		Embedding
		
		& NDCG & HitRate & MRR
		
		& NDCG & HitRate & MRR
		
		& NDCG & HitRate & MRR \\
		
		\midrule
		
		& \multicolumn{9}{c}{@10} \\
		\cmidrule(l{4pt}r{4pt}){2-10}
		
		LUMIR$_{\mathrm{MC}}$
		& 0.005087$^{\ddagger}$ & 0.000327$^{\ddagger}$ & 0.065465
		& 0.009222$^{\ddagger}$ & 0.000028$^{\ddagger}$ & 0.660621
		& 0.084496 & 0.035103$^{\dagger}$ & 0.219529 \\
		
		LUMIR$_{\mathrm{MC}+\mathrm{SA}}$
		& 0.002189$^{\ddagger}$ & 0.000265$^{\ddagger}$ & 0.017876$^{\dagger}$
		& 0.689722 & 0.235593 & 0.613663$^\downarrow$
		& 0.159233$^\downarrow$ & 0.179227$^\downarrow$ & 0.212216$^\downarrow$ \\
		
		\midrule
		
		& \multicolumn{9}{c}{\makebox[0pt]{@20}} \\
		\cmidrule(lr){2-10}
		
		LUMIR$_{\mathrm{MC}}$
		& 0.001152$^{\ddagger}$ & 0.001738$^{\ddagger}$ & 0.017595$^{\dagger}$
		& 0.000070$^{\ddagger}$ & 0.000020$^{\ddagger}$ & 0.191109
		& 0.307402 & 0.929254$^\downarrow$ & 0.325471 \\
		
		LUMIR$_{\mathrm{MC}+\mathrm{SA}}$
		& 0.000091$^{\ddagger}$ & 0.000018$^{\ddagger}$ & 0.005237$^{\ddagger}$
		& 0.202630 & 0.122294 & 0.955319$^\downarrow$
		& 0.013921$^\downarrow$$^{\dagger}$ & 0.001717$^\downarrow$$^{\ddagger}$ & 0.096983$^\downarrow$ \\
		
		\bottomrule
	\end{tabular*}

	{\footnotesize
		$\dagger$: $p < 0.05$, $\ddagger$: $p < 0.01$, based on Welch's t-test comparing each LUMIR variant with the corresponding strongest baseline. 
		
		$\downarrow$ indicates that the corresponding LUMIR variant yields lower performance than the strongest baseline.
	}
\end{table*}

\subsection{Downstream Recommendation Performance}
\paragraph{Overall Performance Comparison.}
The learned POI representations from LUMIR and baseline methods are evaluated on downstream sequential models for next-POI recommendation. Performance is assessed using both accuracy-oriented metrics (HR@K, NDCG@K, MRR@K) and beyond-accuracy metrics (Coverage~\cite{Coverage}, Gini index, Novelty~\cite{Novelty}, and Serendipity~\cite{Serendipity}), with all metrics reported at cutoff values $K \in \{10, 20\}$.

Table~\ref{tab:overall_performance} reports the performance of LUMIR and baseline embedding strategies across three downstream sequential recommendation backbones.\footnote{Due to space constraints, beyond-accuracy results are provided in the online material.}
To examine the impact of the sequential supervision signal $\mathcal{L}_{\mathrm{SA}}$ on embedding quality, an ablation variant is considered in which representations are learned using only the modality compression objective $\mathcal{L}_{\mathrm{MC}}$, without incorporating $\mathcal{L}_{\mathrm{SA}}$.
These two configurations are denoted as LUMIR$_{\mathrm{MC}+\mathrm{SA}}$ and LUMIR$_{\mathrm{MC}}$, respectively. \footnote{We do not include a variant using only $\mathcal{L}_{\mathrm{SA}}$, as this work focuses on modality compression as the primary mechanism for multimodal fusion. The alignment objective alone does not provide sufficient constraints to learn meaningful representations and is treated as an auxiliary component for downstream sequential models.}
The following discussion is organized from the most pronounced performance differences across backbones to the least.

The most pronounced improvements are observed under the BERT4Rec backbone. POI representations learned by LUMIR exhibit consistent and substantial improvements over all baselines across accuracy-oriented metrics. 
Compared to the strongest baseline, LUMIR$_{\mathrm{MC}+\mathrm{SA}}$ achieves improvements of up to +24.2\% in HR@10 and +28.9\% in HR@20. Beyond accuracy, LUMIR also demonstrates clear advantages. It substantially increases coverage while reducing the Gini index, indicating a more balanced recommendation distribution. At the same time, it achieves the highest serendipity scores across both cutoffs, suggesting improved ability to recommend relevant yet unexpected POIs. These results indicate that the joint objective enhances both predictive performance and the diversity-related properties of the learned representations.

Moderate but consistent improvements are observed under the SASRec backbone. LUMIR remains competitive with, and in several cases superior to, baseline methods. In particular, LUMIR$_{\mathrm{MC}}$ consistently delivers the strongest performance, indicating that modality compression alone is sufficient to yield substantial improvements in this setting. In terms of accuracy-oriented metrics, LUMIR$_{\mathrm{MC}}$ achieves the highest performance across most metrics, with notable gains such as +24.5\% in HR@10 and +26.5\% in HR@20 compared to the strongest baseline. In contrast, incorporating the sequence-aware alignment objective does not lead to equally pronounced improvements, suggesting that its contribution is less dominant under this backbone. For beyond-accuracy metrics, LUMIR continues to demonstrate clear advantages. 
LUMIR variants achieve the best results across all metrics, improving both predictive accuracy and diversity-related properties.

In contrast, the differences become less pronounced under the GRU4Rec backbone. LUMIR$_{\mathrm{MC}}$ continues to provide consistent improvements over most baselines across accuracy-oriented metrics. Except for HitRate@20, where it remains comparable to the strongest baseline, LUMIR$_{\mathrm{MC}}$ achieves higher performance on all other metrics, with gains such as +10.3\% in NDCG@10, +11.6\% in HR@10, and +9.2\% in MRR@10. Compared to LUMIR$_{\mathrm{MC}}$, incorporating the sequence-aware alignment objective does not yield additional benefits for GRU4Rec. For beyond-accuracy metrics, LUMIR$_{\mathrm{MC}}$ achieves the best performance on key indicators such as serendipity and novelty, but the overall advantage is less pronounced compared to BERT4Rec and SASRec.

\paragraph{Statistical Significance Analysis.}
To assess whether the observed performance improvements are statistically reliable, statistical significance is evaluated based on per-user evaluation results. Specifically, for each backbone and accuracy-oriented metric, performance distributions are collected across multiple trials, and Welch’s t-test is applied to compare each LUMIR variant against the strongest baseline.

As reported in Table~\ref{tab:significance}, the improvements achieved by LUMIR are statistically significant under the BERT4Rec backbone across most accuracy-oriented metrics, with $p < 0.01$ in the majority of cases. This confirms that the substantial gains observed in Table~\ref{tab:overall_performance} are not due to random variation but reflect consistent improvements. 

Under SASRec and GRU4Rec, the statistical evidence is comparatively weaker, with fewer metrics reaching significance thresholds than in BERT4Rec. Nevertheless, LUMIR continues to achieve statistically significant improvements on several metrics, particularly for the modality compression variant, indicating that the gains remain consistent across instances.

\paragraph{Cross-Backbone Analysis.}
The performance of LUMIR varies across downstream backbones. 
While LUMIR$_{\mathrm{MC}+\mathrm{SA}}$ improves performance under BERT4Rec, its impact is weaker or less consistent under SASRec and GRU4Rec.
The reasons for these differences across backbones are as follows.

This discrepancy can be attributed to the interaction between the alignment objective and the inductive biases of different sequential models. BERT4Rec adopts a bidirectional self-attention mechanism, where each position is conditioned on both past and future context~\cite{sun2019bert4rec}. In this setting, the sequence-aware alignment objective is naturally compatible, as it encourages representations to capture global relational dependencies within the sequence. 
In contrast, SASRec and GRU4Rec follow an autoregressive paradigm, where predictions are strictly conditioned on past interactions~\cite{kang2018self, hidasi2015session}. Under such unidirectional modeling, the contribution of the alignment objective becomes less dominant, as these models primarily rely on directional sequential dependencies. Nevertheless, the learned representations remain effective, and the modality compression objective alone already provides strong performance gains. These observations suggest that LUMIR can adapt to different sequential architectures, while the alignment objective yields the greatest benefits when global contextual modeling is supported.

\section{Conclusion and Future Work}
\textcolor{black}{This work proposes LUMIR, a decoupled multimodal representation learning framework that learns unified POI representations from multimodal tourism-related information collected from heterogeneous data sources by leveraging LLM-based semantic unification and contrastive learning.} By separating representation learning from downstream objectives, LUMIR captures intrinsic cross-modal interactions while incorporating sequence-aware signals as a weak structural constraint. Experimental results across multiple sequential recommendation backbones show that LUMIR consistently improves both accuracy-oriented and beyond-accuracy metrics, demonstrating the effectiveness of decoupled multimodal representation learning from heterogeneous data sources. 

\textcolor{black}{Future work will explore more effective representation learning strategies for multimodal information collected from heterogeneous data sources, particularly in the presence of noisy, redundant, or only partially informative modality-specific signals.}
\textcolor{black}{Specifically, new learning objectives could be developed to better distinguish informative cross-modal interactions from noise while capturing redundancy, complementarity, and semantic consistency across multimodal information.}
\textcolor{black}{Extending LUMIR to more diverse multimodal recommendation scenarios, such as user-centric recommendation and cross-domain applications, will enable further evaluate of the robustness and transferability of the learned representations across heterogeneous data sources.}

\newpage
\section*{Declaration on Generative AI}  
During the preparation of this work, the authors used ChatGPT in order to perform grammar and spelling checks and minor language polishing. After using this tool, the authors reviewed and edited the content as needed and take full responsibility for the publication’s content.

\bibliography{references}


\end{document}


\appendix
\section{Additional Beyond-accuracy Results on the Yelp Dataset}
\begin{table}[tbh]
	
	\centering
	
	\footnotesize
	
	\setlength{\tabcolsep}{3pt}
	
	\renewcommand{\arraystretch}{0.84}
	
	\caption{Performance with different sequential recommendation backbones}
	
	\label{tab:appendix_beyond}
	

	\begin{subtable}{\textwidth}

		\centering
		
		\caption{BERT4Rec Backbone}

		\begin{tabular*}{\textwidth}{@{\extracolsep{\fill}}
						>{\centering\arraybackslash}m{0.8cm}   %
						>{\raggedright\arraybackslash}p{2cm}
						l l l l
					}
				\midrule
				
				\multirow{17}{*}{
						\parbox[c][6cm][c]{1cm}{
								\centering
								\rotatebox[origin=c]{90}{\textit{Beyond-accuracy}}
							}
					}
				& \multicolumn{1}{c}{Embedding}
				& \multicolumn{1}{c}{Coverage}
				& \multicolumn{1}{c}{Gini}
				& \multicolumn{1}{c}{Serendipity}
				& \multicolumn{1}{c}{Novelty} \\
				
				\cmidrule(l{4pt}r{0pt}){2-6}   %
				
				& &\multicolumn{4}{c}{\textit{@10}} \\
				\cmidrule(l{4pt}r{0pt}){3-6}
				\addlinespace[0.2em]
				
				&Trajectory-Only
				& \underline{0.08957$\pm$0.013}
				& \underline{0.9657$\pm$0.0051}
				& 0.0377$\pm$0.0049
				& \underline{\textbf{13.26}$\pm$0.10} \\	[0.3em]
				
				&Full Concatenation
				& 0.02292$\pm$0.0046
				& 0.9914$\pm$0.0013
				& 0.04073$\pm$0.0042
				& 12.23$\pm$0.37 \\ [0.3em]
				
				&Modality-Weighted
				& 0.04937$\pm$0.018
				& 0.9825$\pm$0.0065
				& \underline{0.04716$\pm$0.0023}
				& 12.69$\pm$0.43 \\ [0.3em]
				
				&FLAVA-Based
				& 0.03321$\pm$0.018
				& 0.9874$\pm$0.0065
				& 0.04417$\pm$0.0019
				& 12.49$\pm$0.43 \\ [0.3em]
				
				&CoMM-Based
				& 0.03319$\pm$0.013
				& 0.9874$\pm$0.0049
				& 0.04354$\pm$0.0015
				& 12.76$\pm$0.20 \\ [0.3em]
				
				&LUMIR$_{\mathrm{MC}}$
				& 0.1253$\pm$0.012
				& 0.9516$\pm$0.0055
				& 0.05196$\pm$0.0021
				& 12.83$\pm$0.33 \\ [0.3em]
				
				&LUMIR$_{\mathrm{MC}+\mathrm{SA}}$
				& \textbf{0.1515}$\pm$0.019
				& \textbf{0.9385}$\pm$0.010
				& \textbf{0.05739}$\pm$0.0057
				& 12.88$\pm$0.48 \\

				&& \multicolumn{4}{c}{\textit{@20}} \\
				\cmidrule(lr){3-6}
				\addlinespace[0.2em]
				
				&Trajectory-Only
				& \underline{0.1431$\pm$0.019}
				& \underline{0.9454$\pm$0.0074}
				& 0.06236$\pm$0.0058
				& \underline{\textbf{13.29}$\pm$0.10} \\ [0.3em]
				
				&Full Concatenation
				& 0.03882$\pm$0.0072
				& 0.9845$\pm$0.0023
				& 0.06191$\pm$0.0055
				& 12.32$\pm$0.37 \\ [0.3em]
				
				&Modality-Weighted
				& 0.08263$\pm$0.029
				& 0.9696$\pm$0.010
				& \underline{0.07205$\pm$0.0080}
				& 12.79$\pm$0.39 \\ [0.3em]
				
				&FLAVA-Based
				& 0.05606$\pm$0.029
				& 0.9774$\pm$0.011
				& 0.06434$\pm$0.0036
				& 12.57$\pm$0.42 \\ [0.3em]
				
				&CoMM-Based
				& 0.05588$\pm$0.019
				& 0.9776$\pm$0.0077
				& 0.06571$\pm$0.0041
				& 12.83$\pm$0.18 \\ [0.3em]
				
				&LUMIR$_{\mathrm{MC}}$
				& 0.2004$\pm$0.017
				& 0.9237$\pm$0.0080
				& 0.08526$\pm$0.0054
				& 12.90$\pm$0.29 \\ [0.3em]
				
				&LUMIR$_{\mathrm{MC}+\mathrm{SA}}$
				& \textbf{0.2408}$\pm$0.025
				& \textbf{0.9057}$\pm$0.014
				& \textbf{0.09377}$\pm$0.0062
				& 12.95$\pm$0.41 \\
				\bottomrule
			\end{tabular*}
	
	\end{subtable}
	
		\vspace{1em}
	\begin{subtable}{\textwidth}
		
		\centering
		
		\caption{SASRec Backbone}
\begin{tabular*}{\textwidth}{@{\extracolsep{\fill}}
		>{\centering\arraybackslash}m{0.8cm}   %
		>{\raggedright\arraybackslash}p{2cm}
		l l l l
	}
	\midrule
	
	\multirow{17}{*}{
		\parbox[c][6cm][c]{1cm}{
			\centering
			\rotatebox[origin=c]{90}{\textit{Beyond-accuracy}}
		}
	}
	& \multicolumn{1}{c}{Embedding}
	& \multicolumn{1}{c}{Coverage}
	& \multicolumn{1}{c}{Gini}
	& \multicolumn{1}{c}{Serendipity}
	& \multicolumn{1}{c}{Novelty} \\
	
	\cmidrule(l{4pt}r{0pt}){2-6}   %
	
	& &\multicolumn{4}{c}{\textit{@10}} \\
	\cmidrule(l{4pt}r{0pt}){3-6}
	\addlinespace[0.2em]
	
	&Trajectory-Only
	& 0.02308$\pm$0.013
	& 0.9915$\pm$0.0035
	& 0.02969$\pm$0.0051
	& 11.58$\pm$0.65 \\ [0.3em]
	
	&Full Concatenation
	& 0.04858$\pm$0.0073
	& 0.9833$\pm$0.0029
	& 0.03277$\pm$0.0045
	& \underline{12.61$\pm$0.18} \\ [0.3em]
	
	&Modality-Weighted
	& 0.04860$\pm$0.011
	& 0.9849$\pm$0.0042
	& \underline{0.04064$\pm$0.0028}
	& 12.52$\pm$0.21 \\ [0.3em]
	
	&FLAVA-Based
	& \underline{0.04901$\pm$0.0097}
	& \underline{0.9830$\pm$0.0035}
	& 0.04019$\pm$0.0025
	& 12.23$\pm$0.14 \\ [0.3em]
	
	&CoMM-Based
	& 0.04633$\pm$0.016
	& 0.9836$\pm$0.0048
	& 0.03313$\pm$0.0059
	& 12.52$\pm$0.47 \\ [0.3em]
	
	&LUMIR$_{\mathrm{MC}}$
	& 0.1013$\pm$0.010
	& 0.9658$\pm$0.0052
	&  \textbf{0.05114}$\pm$0.0038
	&  \textbf{12.73}$\pm$0.14 \\ [0.3em]
	
	&LUMIR $_{\mathrm{MC}+\mathrm{SA}}$
	& \textbf{0.1434}$\pm$0.010
	& \textbf{0.9467}$\pm$0.0050
	& 0.04327$\pm$0.0062
	& 12.59$\pm$0.45 \\
	
	&& \multicolumn{4}{c}{\textit{@20}} \\
	\cmidrule(lr){3-6}
	\addlinespace[0.2em]
	
	&Trajectory-Only
	& 0.03916$\pm$0.022
	& 0.9847$\pm$0.0062
	& 0.04779$\pm$0.0084
	& 11.76$\pm$0.60 \\ [0.3em]
	
	&Full Concatenation
	& 0.07992$\pm$0.011
	& 0.9719$\pm$0.0042
	& 0.05413$\pm$0.0071
	& \underline{12.70$\pm$0.16} \\ [0.3em]
	
	&Modality-Weighted
	& \underline{0.08107$\pm$0.018}
	& 0.9746$\pm$0.0068
	& \underline{0.06119$\pm$0.0059}
	& 12.65$\pm$0.21 \\ [0.3em]
	
	&FLAVA-Based
	& 0.08042$\pm$0.016
	& \underline{0.9717$\pm$0.0055}
	& 0.05947$\pm$0.0045
	& 12.39$\pm$0.12 \\ [0.3em]
	
	&CoMM-Based
	& 0.07621$\pm$0.025
	& 0.9719$\pm$0.0076
	& 0.05367$\pm$0.0066
	& 12.63$\pm$0.43 \\ [0.3em]
	
	&LUMIR$_{\mathrm{MC}}$
	& 0.1607$\pm$0.014
	& 0.9470$\pm$0.0072
	&  \textbf{0.07974}$\pm$0.0041
	&  \textbf{12.81}$\pm$0.15 \\ [0.3em]
	
	&LUMIR $_{\mathrm{MC}+\mathrm{SA}}$
	& \textbf{0.2315}$\pm$0.018
	& \textbf{0.9161}$\pm$0.0076
	& 0.06942$\pm$0.0083
	& 12.74$\pm$0.39 \\
	\bottomrule
\end{tabular*}
			
	\end{subtable}
	
			\vspace{1em}
	
	\begin{subtable}{\textwidth}
		
		\centering
		
		\caption{GRU4Rec Backbone}

\begin{tabular*}{\textwidth}{@{\extracolsep{\fill}}
		>{\centering\arraybackslash}m{0.8cm}   %
		>{\raggedright\arraybackslash}p{2cm}
		l l l l
	}
	\midrule
	
	\multirow{17}{*}{
		\parbox[c][6cm][c]{1cm}{
			\centering
			\rotatebox[origin=c]{90}{\textit{Beyond-accuracy}}
		}
	}
	& \multicolumn{1}{c}{Embedding}
	& \multicolumn{1}{c}{Coverage}
	& \multicolumn{1}{c}{Gini}
	& \multicolumn{1}{c}{Serendipity}
	& \multicolumn{1}{c}{Novelty} \\
	
	\cmidrule(l{4pt}r{0pt}){2-6}   %
	
	& &\multicolumn{4}{c}{\textit{@10}} \\
	\cmidrule(l{4pt}r{0pt}){3-6}
	\addlinespace[0.2em]
	
	&Trajectory-Only
	& \underline{\textbf{0.06417}$\pm$0.013}
	& \underline{\textbf{0.9826}$\pm$0.0031}
	& 0.03050$\pm$0.0039
	& 11.62$\pm$0.11 \\ [0.3em]
	
	&Full Concatenation
	& 0.01215$\pm$0.0047
	& 0.9949$\pm$0.0016
	& 0.02552$\pm$0.0016
	& 11.34$\pm$0.21 \\ [0.3em]
	
	&Modality-Weighted
	& 0.04122$\pm$0.012
	& 0.9848$\pm$0.0046
	& 0.03358$\pm$0.0057
	& \underline{11.99$\pm$0.29} \\ [0.3em]
	
	&FLAVA-Based
	& 0.02857$\pm$0.0089
	& 0.9889$\pm$0.0035
	& \underline{0.03394$\pm$0.0046}
	& 11.77$\pm$0.24 \\ [0.3em]
	
	&CoMM-Based
	& 0.02786$\pm$0.0057
	& 0.9893$\pm$0.0021
	& 0.03114$\pm$0.0036
	& 11.78$\pm$0.14 \\ [0.3em]
	
	&LUMIR$_{\mathrm{MC}}$
	& 0.04398$\pm$0.019
	& 0.9833$\pm$0.0067
	& \textbf{0.03892}$\pm$0.0045
	& \textbf{12.14}$\pm$0.44 \\ [0.3em]
	
	&LUMIR$_{\mathrm{MC}+\mathrm{SA}}$
	& 0.02653$\pm$0.011
	& 0.9898$\pm$0.0039
	& 0.03105$\pm$0.0045
	& 11.73$\pm$0.38 \\
	\addlinespace[0.2em]
	
	&& \multicolumn{4}{c}{\textit{@20}} \\
	\cmidrule(lr){3-6}
	\addlinespace[0.2em]
	
	&Trajectory-Only
	& \underline{\textbf{0.1081}$\pm$0.027}
	& \underline{0.9716$\pm$0.0048}
	& 0.04499$\pm$0.0040
	& 11.83$\pm$0.095 \\ [0.3em]
	
	&Full Concatenation
	& 0.02090$\pm$0.0080
	& 0.9905$\pm$0.0027
	& 0.03856$\pm$0.0029
	& 11.55$\pm$0.18 \\ [0.3em]
	
	&Modality-Weighted
	& 0.06947$\pm$0.021
	& 0.9735$\pm$0.0077
	& \underline{\textbf{0.05621}$\pm$0.0060}
	& \underline{12.16$\pm$0.27} \\ [0.3em]
	
	&FLAVA-Based
	& 0.04804$\pm$0.015
	& 0.9804$\pm$0.0059
	& 0.05377$\pm$0.0050
	& 11.95$\pm$0.22 \\ [0.3em]
	
	&CoMM-Based
	& 0.04705$\pm$0.0093
	& 0.9809$\pm$0.0034
	& 0.04915$\pm$0.0044
	& 11.96$\pm$0.12 \\ [0.3em]
	
	&LUMIR$_{\mathrm{MC}}$
	& 0.07264$\pm$0.030
	& \textbf{0.9715}$\pm$0.011
	& 0.05567$\pm$0.0047
	& \textbf{12.28}$\pm$0.39 \\ [0.3em]
	
	&LUMIR$_{\mathrm{MC}+\mathrm{SA}}$
	& 0.04502$\pm$0.018
	& 0.9821$\pm$0.0065
	& 0.04526$\pm$0.0064
	& 11.91$\pm$0.35 \\
	\bottomrule
\end{tabular*}

	\end{subtable}
	
	{\footnotesize
	\textbf{Bold} and \underline{underline} indicate the highest mean value for each metric and the best-performing baseline, respectively. Results are averaged over 8 runs after removing the highest and lowest from 10 independent runs.
}
\end{table}